\documentclass[cernpreprint, atlasdraft=false, texlive=2023, UKenglish, texmf, orcidlogo]{atlasdoc}
 
\usepackage{atlaspackage}
\usepackage{atlasbiblatex}
 
\usepackage{atlasphysics, comment,adjustbox}

\graphicspath{{logos/}{figures/}}
 
\usepackage{ANA-SUSY-2020-09-PAPER-defs}

\AtlasTitle{Search for pairs of muons with small displacements in $pp$ collisions at $\sqrt{s} = 13$ TeV with the ATLAS detector}

\AtlasAbstract{
A search for new phenomena giving rise to pairs of opposite electrically charged muons with impact parameters in the millimetre range is presented, using $139$ fb$^{-1}$ of $\sqrt{s} = 13$ TeV $pp$ collision data from the ATLAS detector at the LHC. The search targets the gap in coverage between existing searches targeting final states with leptons with large displacement and prompt leptons. No significant excess over the background expectation is observed and exclusion limits are set on the mass of long-lived scalar supersymmetric muon-partners (smuons) with much lower lifetimes than previously targeted by displaced muon searches. Smuon lifetimes down to 1 ps are excluded for a smuon mass of 100 GeV, and smuon masses up to 520 GeV are excluded for a proper lifetime of 10 ps, at 95\% confidence level. Finally, model-independent limits are set on the contribution from new phenomena to the signal-region yields.
}
 
\author{The ATLAS Collaboration}

\PreprintIdNumber{CERN-EP-2023-038}

\AtlasJournal{Phys.Lett. B}
\AtlasJournalRef{\PLB 846 (2023) 138172}
\AtlasDOI{10.1016/j.physletb.2023.138172}

\hypersetup{pdftitle={ATLAS document},pdfauthor={The ATLAS Collaboration}}
 
\begin{document}
 
\maketitle

\section{Introduction}

The Standard Model (SM) contains many particles that have a significant lifetime which, when produced at a collider, travel a certain distance before decaying away from the primary proton-proton (\pp) interaction. Despite this, the majority of beyond the standard model (BSM) searches at the Large Hadron Collider (LHC) focus on prompt decays, and are not optimised for particles that travel a macroscopic distance. 
Many BSM theories predict particles that have significant lifetimes including $R$-parity-conserving supersymmetry (SUSY)~\cite{Farrar:1978xj,Golfand:1971iw,Volkov:1973ix,Wess:1974tw,Wess:1974jb,Ferrara:1974pu,Salam:1974ig} as well as $R$-parity-violating SUSY models~\cite{Barbier:2004ez,Allanach:2006st}, models like split-SUSY~\cite{Giudice:2004tc,ArkaniHamed:2004fb}, exotic scenarios such as universal extra dimensions~\cite{Appelquist:2000nn,Cheng:2002ab} and gauge-mediated SUSY breaking (GMSB)~\cite{Dine:1981gu,AlvarezGaume:1981wy,Nappi:1982hm}.
 
In GMSB SUSY models the lightest SUSY particle (LSP) is a nearly massless gravitino, and the next-to-lightest SUSY particle (NLSP) becomes long-lived due to the small coupling to the LSP.
Well-motivated versions of this model have a stau ($\tilde{\tau}$) as the single NLSP, or a selectron ($\tilde{e}$), smuon ($\tilde{\mu}$), and $\tilde{\tau}$ as a set of degenerate co-NLSPs~\cite{Evans_2016}. In these models, pair-produced sleptons ($\tilde{\ell}$) of the same flavour decay into an invisible gravitino and a charged lepton of the same flavour as the parent $\tilde{\ell}$.
 
\begin{figure}[th]
\centering
\includegraphics[width=0.5\textwidth]{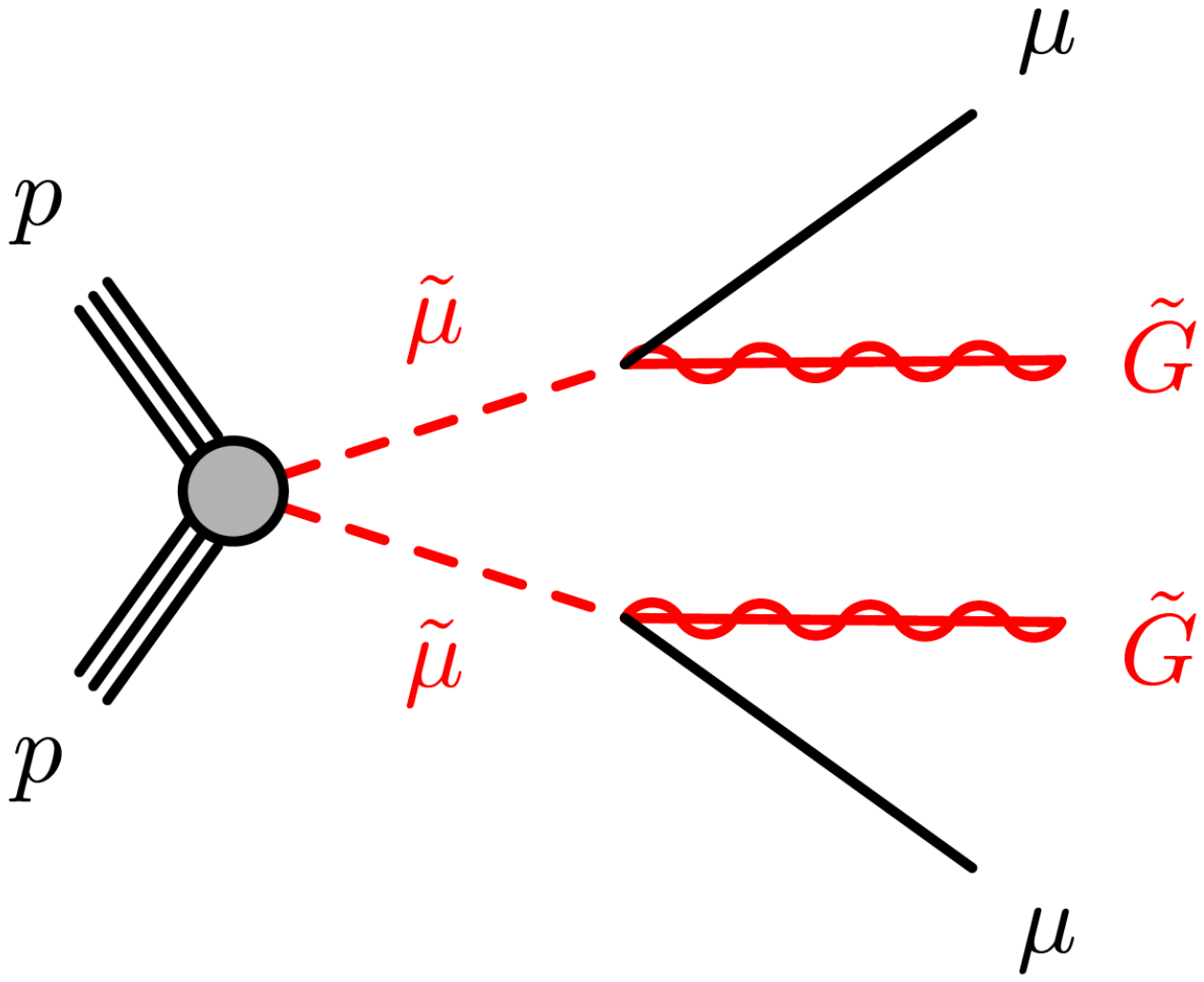}
\caption{Decay topology of the simplified model considered where smuons ($\tilde{\mu}$) are pair produced and each smuon decays to a gravitino ($\tilde{G}$) and a muon ($\mu$).}
\label{fig:feynman}
\end{figure}

This paper presents a search for supersymmetric partners of the muon ($\tilde{\mu}$) with a lifetime of $\mathcal{O}(1-10)$~ps, targeting a gap in coverage between prompt slepton searches, and displaced slepton searches which have optimal sensitivity for lifetimes around $\mathcal{O}(100-1000)\text{ ps}$. This regime has been highlighted as a possible blind spot in BSM searches at the LHC~\cite{Heinemeyer:2021opc}. Figure~\ref{fig:feynman} shows a diagram of the targeted signal.
A combination of results from the LEP experiments excludes the superpartners of the right-handed muons ($\tilde{\mu}_R$) of any lifetime for masses less than 96.3~\GeV~\cite{Heister:2002vh,Abbiendi:2005gc,L3,Abdallah:2002rd,LEPlimitsSlepton}. Previous searches for long-lived sleptons have been performed by the ATLAS~\cite{SUSY-2018-14} and CMS~\cite{CMS-EXO-18-003} collaborations, excluding smuons up to 700~\GeV\ and 620~\GeV\ respectively, for a lifetime of 100~ps.
 
\section{ATLAS detector}

\newcommand{\AtlasCoordFootnote}{
ATLAS uses a right-handed coordinate system with its origin at the nominal interaction point (IP)
in the centre of the detector and the \(z\)-axis along the beam pipe.
The \(x\)-axis points from the IP to the centre of the LHC ring,
and the \(y\)-axis points upwards.
Cylindrical coordinates \((r,\phi)\) are used in the transverse plane,
\(\phi\) being the azimuthal angle around the \(z\)-axis.
The pseudorapidity is defined in terms of the polar angle \(\theta\) as \(\eta = -\ln \tan(\theta/2)\).
Angular distance is measured in units of \(\Delta R \equiv \sqrt{(\Delta\eta)^{2} + (\Delta\phi)^{2}}\).}

The ATLAS detector~\cite{PERF-2007-01} at the LHC covers nearly the entire solid angle around the collision point.\footnote{\AtlasCoordFootnote}
It consists of an inner tracking detector surrounded by a thin superconducting solenoid, electromagnetic and hadron calorimeters,
and a muon spectrometer incorporating three large superconducting air-core toroidal magnets.
 
The inner-detector system~(ID) is immersed in a \SI{2}{\tesla} axial magnetic field
and provides charged-particle tracking in the range \(|\eta| < 2.5\).
The high-granularity silicon pixel detector covers the vertex region and typically provides four measurements per track,
the first hit normally being in the insertable B-layer installed before Run~2~\cite{ATLAS-TDR-19,PIX-2018-001}.
It is followed by the silicon microstrip tracker, which usually provides eight measurements per track.
These silicon detectors are complemented by the transition radiation tracker (TRT),
which enables radially extended track reconstruction up to \(|\eta| = 2.0\).
The TRT also provides electron identification information
based on the fraction of hits (typically 30 in total) above a higher energy-deposit threshold corresponding to transition radiation.
 
The calorimeter system covers the pseudorapidity range \(|\eta| < 4.9\).
Within the region \(|\eta|< 3.2\), electromagnetic calorimetry is provided by barrel and
endcap high-granularity lead/liquid-argon~(LAr) calorimeters,
with an additional thin LAr presampler covering \(|\eta| < 1.8\)
to correct for energy loss in material upstream of the calorimeters.
Hadron calorimetry is provided by the steel/scintillator-tile calorimeter,
segmented into three barrel structures within \(|\eta| < 1.7\), and two copper/LAr hadron endcap calorimeters.
The solid angle coverage is completed with forward copper/LAr and tungsten/LAr calorimeter modules
optimised for electromagnetic and hadronic energy measurements respectively.
 
The muon spectrometer~(MS) comprises separate trigger and
high-precision tracking chambers measuring the deflection of muons in a magnetic field generated by the superconducting air-core toroidal magnets.
The field integral of the toroids ranges between \num{2.0} and \SI{6.0}{\tesla\metre}
across most of the detector.
Three layers of precision chambers, each consisting of layers of monitored drift tubes, covers the region \(|\eta| < 2.7\),
complemented by cathode-strip chambers in the forward region, where the background is highest.
The muon trigger system covers the range \(|\eta| < 2.4\) with resistive-plate chambers in the barrel, and thin-gap chambers in the endcap regions.
 
Interesting events are selected by the first-level trigger system implemented in custom hardware,
followed by selections made by algorithms implemented in software in the high-level trigger~\cite{TRIG-2016-01}.
The first-level trigger accepts events from the \SI{40}{\MHz} bunch crossings at a rate below \SI{100}{\kHz},
which the high-level trigger further reduces in order to record events to disk at about \SI{1}{\kHz}.
 
An extensive software suite~\cite{ATL-SOFT-PUB-2021-001} is used in the reconstruction and analysis of real
and simulated data, in detector operations, and in the trigger and data acquisition systems of the experiment.


\section{Data and simulated event samples}

The data set used in this analysis was collected by the ATLAS detector in \pp\ collisions provided by the LHC during Run~2 from 2015 to 2018. The beams collided at a centre-of-mass energy of $\sqrt{s} = 13$~\TeV\ and with a minimum separation of 25~ns between consecutive proton bunch  crossings. The average number $\langle \mu \rangle$ of additional \pp\ interactions per bunch crossing (pile-up) ranged from 14 in 2015 to about 38 in 2017--2018. After data-quality requirements~\cite{DAPR-2018-01}, applied to ensure that all parts of the detector were operational during data-taking, the data sample amounts to a total integrated luminosity of $139~\ifb$. Candidate events were selected by a di-muon trigger~\cite{TRIG-2016-01,TRIG-2018-01}, where the trigger has no explicit selection cuts on the transverse impact parameter of the muon's track with respect to the beam line, \dzero.
 
To evaluate signal sensitivity, Monte Carlo (MC) events in a simplified GMSB SUSY model were simulated with up to two additional partons at leading order using \MGMCatNLOV{v2.6.1}~\cite{Alwall:2014hca} with the \nnpdftwo parton distribution function (PDF) set~\cite{Ball:2012cx}, and interfaced to \PYTHIAV{8.230}~\cite{Sjostrand:2006za} using the A14 set of tuned parameters~\cite{ATL-PHYS-PUB-2014-021}. The smuon decay was simulated using $\GEANT$~\cite{Agostinelli:2002hh}.
The impact of multiple interactions in the same and neighbouring bunch crossings (pileup) was modeled by overlaying each hard-scattering event with simulated minimum-bias events generated with \PYTHIAV{8.210}~\cite{Sjostrand:2006za} using the A3 tune~\cite{ATL-PHYS-PUB-2016-017} and \nnpdftwo PDF set~\cite{Ball:2012cx}. Signal cross sections were calculated at next-to-leading order in $\alpha_\text{s}$, with soft-gluon emission effects added at next-to-leading-logarithm accuracy ~\cite{Beenakker:1999xh,Debove:2010kf,Fuks:2012qx,Fuks:2013vua,Fiaschi:2018hgm}. The nominal cross section and uncertainty were taken from an envelope of predictions using different PDF sets and factorization and renormalization scales~\cite{Borschensky:2014cia}. The simplified model used for interpretation gives a cross section of $5.4_{- 0.2}^{+0.1}$~pb for a $\tilde{\mu}$ with mass $50~\GeV$ and $0.1 \pm 0.05$~fb for a $\tilde{\mu}$ with mass $700~\GeV$, where the cross section assumes degenerate left- and right-handed smuons. Simulated events were generated for $\tilde{\mu}$ masses ranging from 50--700~\GeV\ and lifetimes of 0.1ps, 1 ps, 10 ps, and 100 ps.
 
The dominant SM background for this search originates from semileptonic $B$-hadron decays, \bbmm. A data driven background method is used to estimate the number of background events in the signal region, described in Section~\ref{sec:bkgest}, and any sub-dominant background processes are included within this method. SM processes with prompt leptons such as $Z$+jets, $W$+jets, $t\overline{t}$, single top quark, and di-boson were assessed with simulation and found to be negligible.

\section{Object reconstruction and event selection}

Candidate events are required to have at least one \pp interaction vertex with a minimum of two associated tracks, where each track has \pt $> 500$ \MeV~\cite{ATL-PHYS-PUB-2015-026}. In events with multiple vertices, the primary vertex is defined as the vertex with the highest scalar sum of the squared transverse momenta of its associated tracks.
 
Muon candidates are reconstructed in the pseudorapidity range $\lvert \eta \rvert < 2.5$ by matching MS tracks with ID tracks. The muons are required to be matched to the muons that fired the trigger. As requirements on the transverse momentum, \pt, of the muons are imposed in the online trigger decision, higher \pt requirements of $\pt > 20$ GeV are applied to the muons to ensure that trigger efficiencies are constant in the relevant phase-space. The muons must satisfy the Medium identification requirements defined in Ref.~\cite{MUON-2018-03}. Nominal track reconstruction was used and it was verified by MC simulations that the nominal tracking reconstruction efficiency, including the muon identification efficiency, is flat as a function of \abdzero in the entire \abdzero region $(0-3)$ mm that is used in this search. Isolation criteria are applied to each muon in order to suppress contributions of muons from the decays of hadronic sources. The isolation variable named ``PflowLoose'' defined in Ref.~\cite{MUON-2018-03} is applied using the working point where the $\Delta R$ parameter is varied.
 
To reduce background contributions from prompt SM processes a selection cut of $\abdzero \geq 0.1$ mm is made. No selection cuts are made on the longitudinal impact parameter, $z_0$.
 
Selected events are required to have at least two muons of opposite electric charge passing all the selection criteria.
Although this search probes non-prompt muons, muons from prompt process, such as from the decay of a $Z$ boson, can be included in the control regions (defined in Section~\ref{sec:bkgest}) due to the resolution of the transverse impact parameter. In order to reduce contributions of di-muon events from the decay of $Z$ bosons the invariant mass of the two muons must be $\geq 110$ \GeV. Further selection cuts are made to provide control regions (CRs), validation regions (VRs) and signal regions (SRs) as described in Section~\ref{sec:bkgest}.


\section{Background estimation}
\label{sec:bkgest}
After selection, the dominant background originates from semileptonic $B$-hadron decays, \bbmm. Possible contributions from processes with prompt leptons such as $Z$+jets, $W$+jets, $t\overline{t}$, single top quark, and di-boson were assessed with MC and were found to be negligible.

An extended ABCD method~\cite{abcd} is used to estimate the number of background events in the signal region. The ABCD method requires two uncorrelated variables, chosen here to be the absolute value of the transverse impact parameter, $\abdzero$, of the muon track for the positively charged muon and the negatively charged muon, to create a two-dimensional plane in data that is split into four sections. The variable \abdzero~is chosen as it is a good proxy for the displacement of muon tracks. To further separate background dominated regions from signal dominated regions the method is extended with a third variable, the invariant mass of the two muons (\invm), which is uncorrelated with both impact parameters to define eight regions, A to H, as shown in Figure~\ref{fig:reg1}. The region H, located at high invariant mass and impact parameter values, is defined as the signal region.
 
Using the ABCD method the number of background events in the signal region H can be estimated by:
\begin{equation}
N^{\text{est. bkg.}}_H = N_A^{\text{data}} \cdot r^{\dzero^+} \cdot  r^{\dzero^-} \cdot  r^{\invm},
\end{equation}
 
where $r^{\dzero^+}$, $r^{\dzero^-}$, and $r^{\invm}$ are ratios from high to low \abdzero of the positive and negative charged muons and \invm regions. The ratios can be computed from multiple pairs of regions and so the regions with highest statistics and lowest signal contamination are used to compute the ratios: $r^{\dzero^+} = N_C^{\text{data}}/N_A^{\text{data}}$, $r^{\dzero^-} = N_B^{\text{data}}/N_A^{\text{data}}$, and $r^{\invm}=N_E^{\text{data}}/N_A^{\text{data}}$, where $N_i^{\text{data}}$ is the number of observed events in data in region $i\in \{A,B,C,E\}$.
 
\begin{figure}[htb]
\centering
\includegraphics[width=0.8\textwidth]{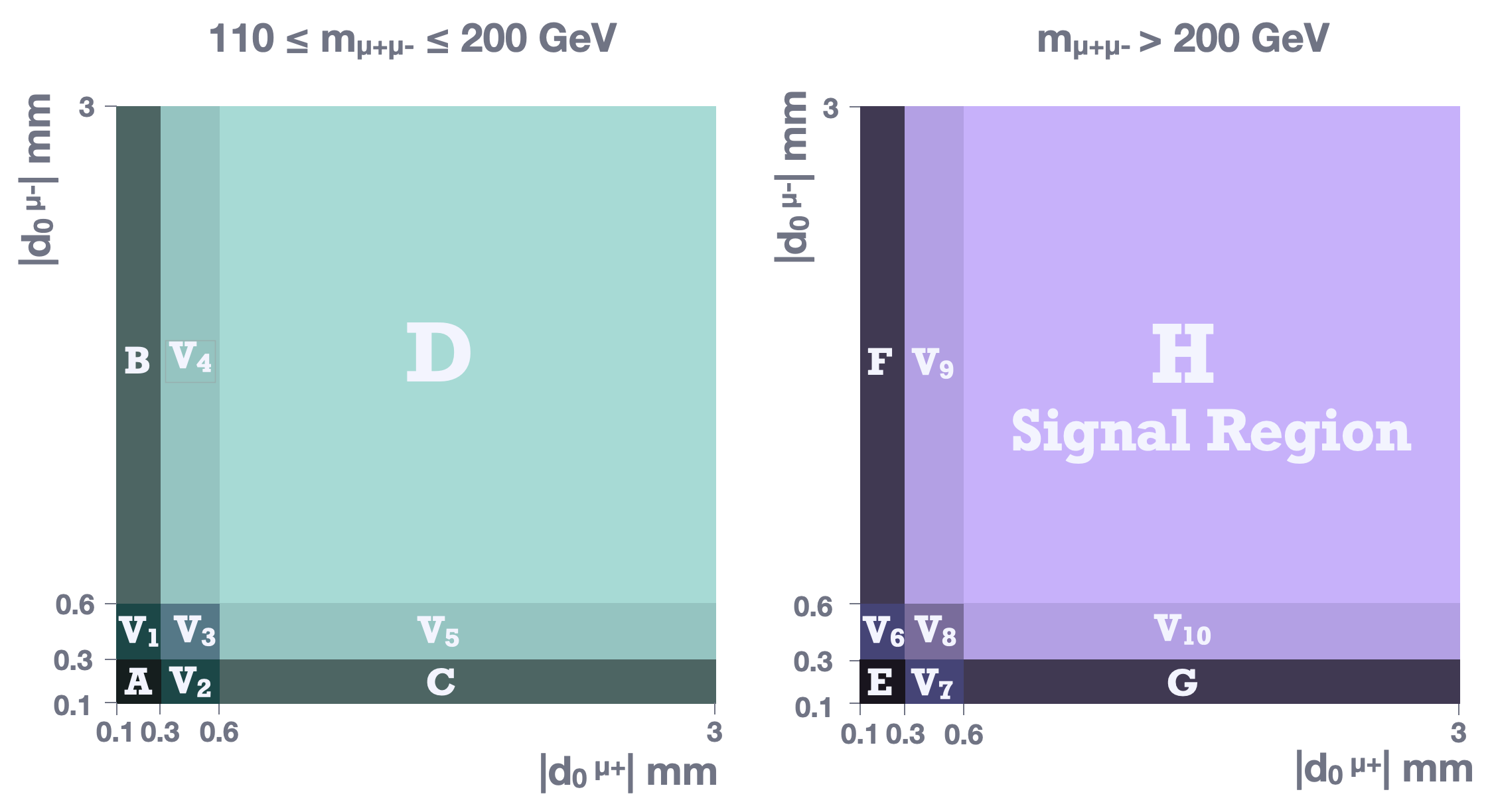}
\caption{A graphic depicting the regions for the extended ABCD background estimation method for Set of Regions 1. Data is split into a 2D plane in \abdzero with the positive charged muon versus the negative charged muon. For the regions on the left the invariant mass of the two muons must be greater or equal to 110 GeV and less than or equal to 200 GeV, the invariant mass of the two muons must be greater than 200 GeV for the regions on the right. A, B, C, and E are control regions. The regions V1-V10, D, F, and G are validation regions, and H is the signal region.}
\label{fig:reg1}
\end{figure}
 
The boundaries between regions are defined to be non-adjacent in $\abdzero$, such that there are ``gap'' regions between each of the control regions (CRs) and signal regions (SRs). The gaps are needed to reduce signal contamination in the CRs, and allow for validation regions (VRs) to be defined to confirm the correctness of the background estimation.
 
Three overlapping sets of regions are defined in Table~\ref{tbl:SRs} and \textit{Set of Regions 1} is depicted in Figure~\ref{fig:reg1}. In the Set of Regions~2 and~3 the invariant mass selection threshold of 200~GeV is lowered to 140 and 125 GeV respectively. An additional selection on the angular distance between the two muons, $\Delta R_{\mu^+\mu^-}$ > 3~radians, is applied in Set of Regions 3. The Set of Regions~1 provides the strongest sensitivity to the GMSB model targeted by this search. However, the Set of Regions~2 and 3 provide two additional SRs, and CRs and VRs without significant contamination from signals with masses $\leq 200$~\GeV, where a potentially large signal contamination in the control regions would bias the background estimation and absorb partly an excess of events in the signal region. These sets of regions are also used to set limits on model-independent BSM signal processes. When performing hypothesis tests only one set of regions is used per signal mass and lifetime point as the three regions are overlapping in events.
 
\begin{table}[th]
\small
\begin{center}
\caption{The definitions of the three sets of regions that are used to define the CRs, VRs, and SRs, where the columns 2 and 3 refer to the boundaries used to subdivide the planes in \abdzero as depicted in Figure~\ref{fig:reg1} for Set of Regions 1.}
\label{tbl:SRs}
\def\arraystretch{1.4}
\begin{tabular}{ c|c|c|c| c }
\toprule
Set of Regions & Lower displacement region  & Higher displacement region & Threshold \invm & Additional cut \\
\hline
1    & $0.1 \leq \abdzero < 0.3$ &$0.6 \leq \abdzero <3$ mm & 200 GeV & - \\
2    & $0.1 \leq \abdzero < 0.3$ &$0.6 \leq \abdzero <3$ mm & 140 GeV & - \\
3    & $0.1 \leq \abdzero < 0.3$ &$0.6 \leq \abdzero <1.3$ mm & 125 GeV & $\Delta R_{\mu^+\mu^-}>3$ rad.\\
\bottomrule
\end{tabular}
\end{center}
\end{table}


The extended ABCD method requires the three variables, \abdzeroplus, \abdzeroneg, and \invm, to be uncorrelated. To quantify potential correlations closure tests are performed in the validation regions V4 to V8 of the three region sets, using regions A, B, C, E, V1, V2 and V3 for the ratios to compute the number of expected events in the validation region. Regions F, G, and V9-V10 are not included in the test due to potential signal contamination. The numbers of estimated and observed events are consistent within statistical uncertainties at the 1$\sigma$ standard deviation level, except for V8 where the standard deviation is found to be $2\sigma$ in Set of Regions 1. Figure~\ref{fig:val} shows expected and observed number of events in the validation regions V4-V8 for each set of regions and the relative difference for each validation region. Based on these results, a conservative uncertainty of 40\% is assigned as non-closure systematic uncertainty in the three signal regions.
 
\begin{figure}[h!]
\centering
\includegraphics[width=0.6\textwidth]{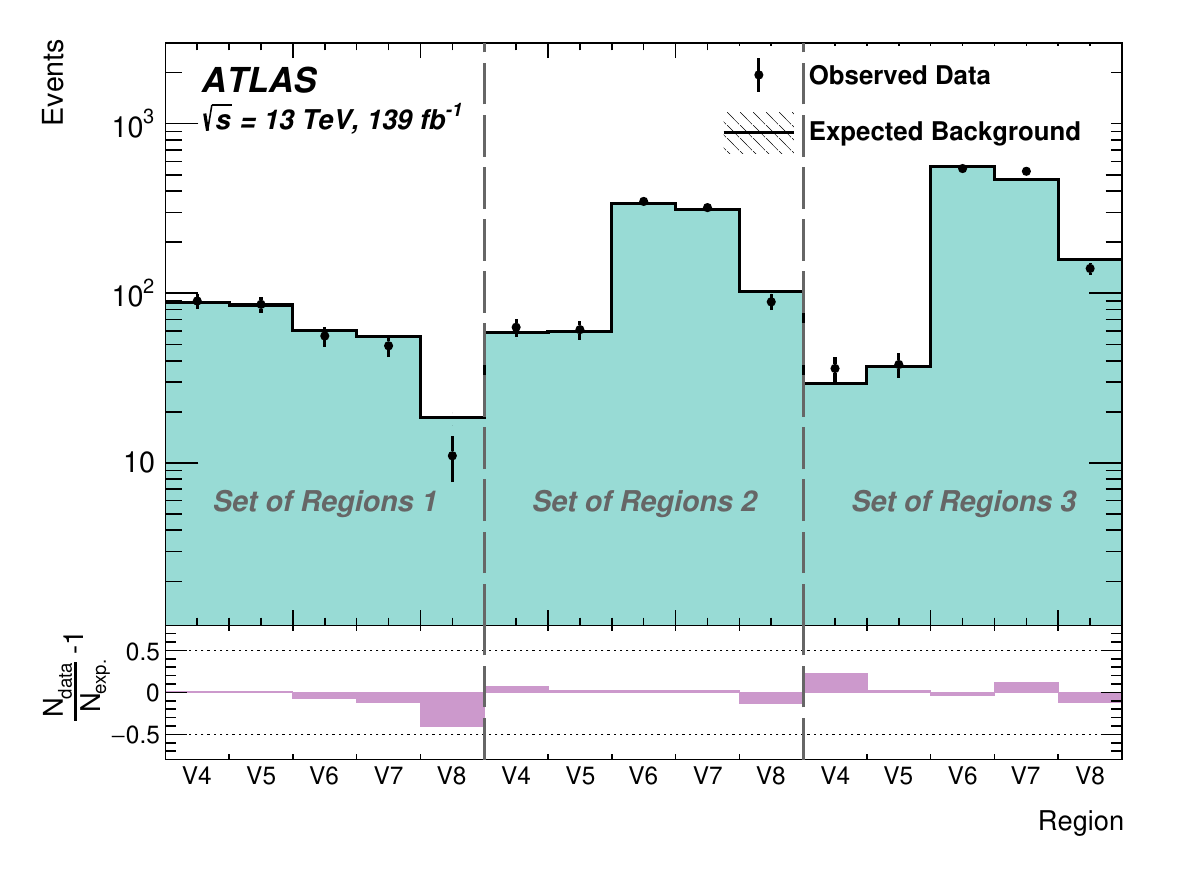}
\caption{
Expected and observed number of events in the validation regions V4-V8 for each set of regions are shown. The ratio panel shows the relative difference for each validation region.
}
\label{fig:val}
\end{figure}


\section{Systematic uncertainties}

Differences in the efficiencies and resolution between data and MC were corrected for by the application of multiplicative factors. Uncertainties on the factor correcting for differences in trigger efficiency are found to be within 5\% for all signal samples and are applied as a systematic uncertainty to the number of expected signal events. The remaining uncertainties that may arise due to the differences between MC and data include muon reconstruction efficiency, the scale and resolution of the momentum of the muon and are found to be negligible, < 1\%, across all analysis regions and no systematic uncertainties are added in these categories. Statistical uncertainties associated with the simulated MC samples are also accounted for. Theoretical uncertainties include cross section uncertainties of 1.6--3.9\%. Modeling uncertainties for the signal process, e.g. effects of varying the factorisation and renormalisation scales, initial and final state radiation, and underlying effects were found to be negligible and are not included as systematics. In order for the exclusion fit to interpolate between the generated lifetimes and extrapolate above the generated lifetimes a re-weighting procedure is applied to the generated signal MC samples to provide smuon lifetimes between the generated lifetimes. Uncertainties on this procedure were evaluated by observing the percentage change when re-weighting the lifetime of a generated sample to that of a different generated lifetime. From these studies a 12\% uncertainty is applied when the lifetime is re-weighted to a higher lifetime. However, when the lifetime is re-weighted to a lower lifetime the agreement between the lifetime of the original and the re-weighted lifetime is well within statistical uncertainty and thus no uncertainty is added in this case.
 
A background uncertainty of $40\%$ for the non-closure of the extended ABCD method is assigned to the estimated number of background events in the signal region.
 
The uncertainty in the combined 2015--2018 integrated luminosity is 1.7\% \cite{ATLAS-CONF-2019-021}, obtained using the LUCID-2 detector \cite{LUCID2} for the primary luminosity measurements.


\section{Results}

The observed number of events in all signal regions, shown in Table~\ref{tab:results}, is found to be compatible with the background expectation within statistical and systematic uncertainties.
 
The results are used to set model-independent limits on the contribution of generic BSM signals in each of the SRs defined by the three sets of regions, assuming no signal contamination in the CRs. Possible signal leakage to the CRs can produce a bias in the background estimation, leading to conservative limits. For the GMSB model the signal contamination for smuon masses of 300, 400 and 500 GeV in Set of Regions 1 is negligible, for Set of Regions 2 the signal contamination is negligible for smuon mass 200 GeV and for the Set of Region 3 the signal contamination is negligible for smuon masses of 50 and 100 GeV. Table~\ref{tab:results} shows the results of a model-independent fit, performed using the HistFitter package~\cite{Baak:2014wma}, where the CL$_\text{s}$ prescription~\cite{Read:2002hq} is used  to set upper limits at 95\% CL on the visible cross-section $\langle A \epsilon{ \sigma}_{obs}^{95}\rangle$, where $A$ is the acceptance and $\epsilon$ the efficiency, as well as on the observed ($S^{95}_{{obs}}$) and expected ($S^{95}_{{exp}}$) number of events from potential new physics processes in the SRs. The $p$-value and the corresponding significance for the background only hypothesis are also evaluated.
 
\begin{table}[htb]
\centering
\caption[Breakdown of upper limits.]{
Presented here are the expected and observed number of background events in each SR H defined by the three sets of regions, where combined statistical and systematic uncertainties on expected $N^{\text{bkg}}_H$ are given. Columns four to six show the 95\% CL upper limits on the visible cross section ($\langle A \epsilon\sigma\rangle_{\text{obs}}^{95}$) and on the number of signal events given the observed and expected number of events ($S_{\text{obs}}^{95}$ and $S_{\text{exp}}^{95}$), where there is a $\pm 1 \sigma$ uncertainty on $S_{\text{exp}}^{95}$. The last two columns indicate the $CL_B$ value, i.e. the confidence level observed for the background-only hypothesis, and the discovery $p$-value ($p(s = 0)$), capped at 0.5.
\label{tab:results}}
\setlength{\tabcolsep}{0.0pc}
\begin{tabular*}{\textwidth}{@{\extracolsep{\fill}}cccccccc}
\noalign{\smallskip}\hline\noalign{\smallskip}
Set of Regions & Expected $N^{\text{bkg}}_H$ & Observed $N^{\text{data}}_{H}$ & $\langle A\epsilon{\sigma}\rangle_{\text{obs}}^{95}$[fb]  &  $S_{\text{obs}}^{95}$  & $S_{\text{exp}}^{95}$ & $CL_{B}$ & $p(s=0)$ ($Z$)  \\
\noalign{\smallskip}\hline\noalign{\smallskip}
1 & $2.1$ $\pm$ $0.8$  & $1$  & $0.02$ &  $3.3$ & $ { 4.2 }^{ +2.5 }_{ -1.4 }$ & $0.27$ & $ 0.50$~$(0.00)$ \\%
 
2 & $12.5$ $\pm$ $5.2$ & $7$ & $0.04$ &  $5.2$ & $ { 8.5 }^{ +4.0 }_{ -2.7 }$ & $0.08$ & $ 0.50$~$(0.00)$ \\%
3 &  $17.2$ $\pm$ $7.4$ & $14$ & $0.06$ &  $8.9$ & $ { 10.5 }^{ +5.0 }_{ -3.1 }$ & $0.26$ & $ 0.50$~$(0.00)$ \\%
\noalign{\smallskip}\hline\noalign{\smallskip}
\end{tabular*}
\end{table}
 
Model-dependent exclusion limits for GMSB SUSY models on the smuon masses and lifetimes are derived at 95\% confidence level following the CL$_\text{s}$ prescription~\cite{Read:2002hq}.
A combined likelihood fit is performed in regions A, B, C, E, and H, including the possible signal contribution in the control regions. The Set of Regions~1 provides the best expected sensitivity across all the plane and therefore is the only set of regions used in the model dependent fit. The HistFitter package~\cite{Baak:2014wma} is used for statistical interpretation, and all systematic uncertainties are treated as Gaussian nuisance parameters during the fitting procedure. A re-weighting procedure is applied to the generated signal samples to provide signal lifetime points between those that are generated. Interpolation is used to provide smooth results throughout the plane, connecting the discrete mass and lifetime values that were simulated and the re-weighted lifetimes.
The results are presented in Figure~\ref{fig:excl_limit1} where smuon lifetimes down to 1~ps and smuon masses up to 520 GeV are excluded, assuming degenerate left- and right-handed smuons.
The results from a previous search for displaced leptons (named here as the Displaced Slepton Signature) with large impact parameter (3 mm $< \abdzero <$ 300 mm) are also shown~\cite{SUSY-2018-14}. A search for direct slepton production with prompt decay~\cite{SUSY-2018-32} is reinterpreted using the RECAST framework~\cite{recast} to cover lifetimes below the picosecond regime (named here as the Prompt Slepton Signature). This is the first explicit reinterpretation of prompt lepton searches in the long-lived regime in ATLAS. The search presented in this paper bridges the gap between both previous searches.
 
\begin{figure}[h!]
\centering
\includegraphics[width=0.6\textwidth]{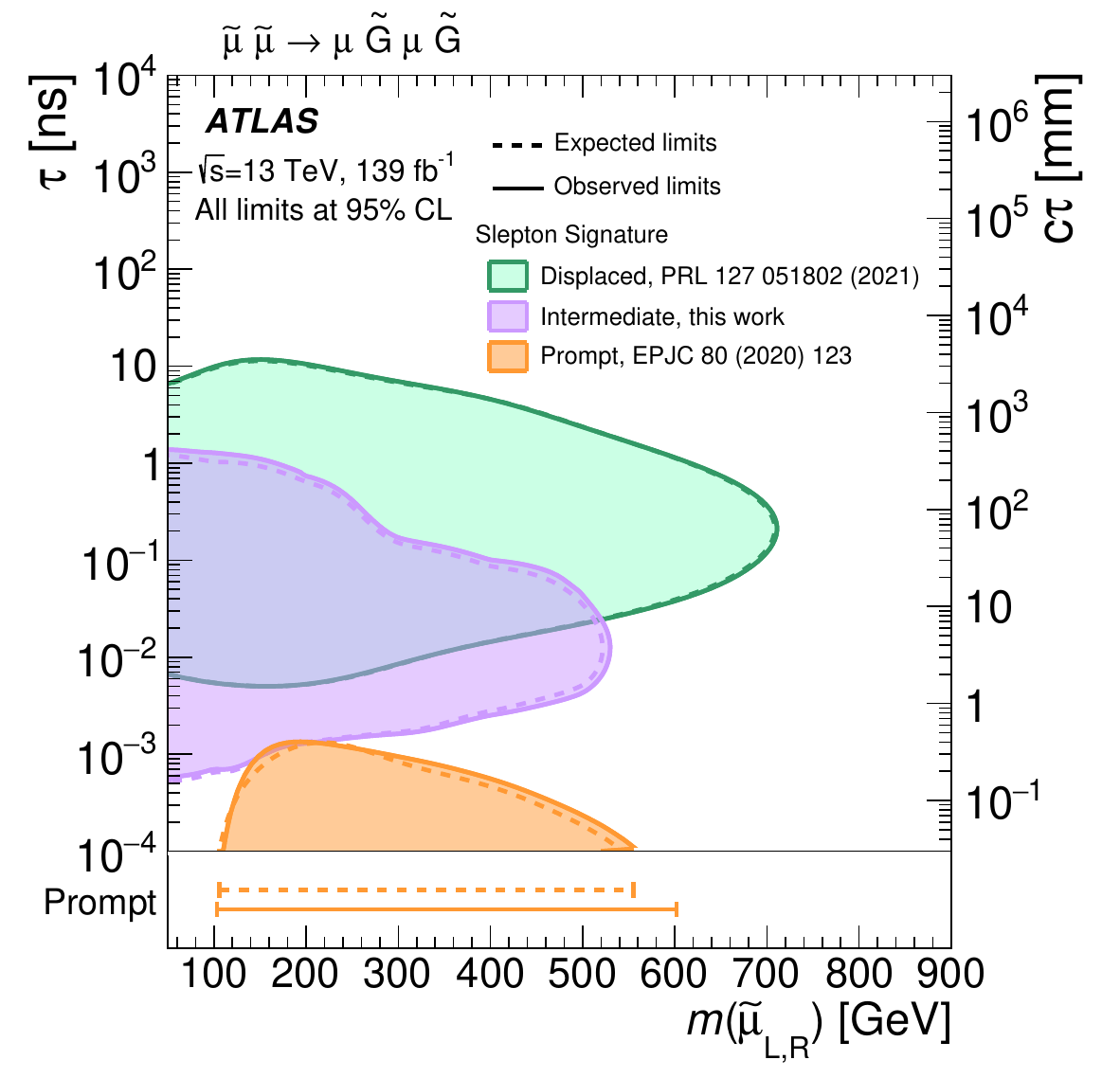}
\caption{
Expected (dashed) and observed (solid) exclusion contours for $\tilde{\mu}$ NLSP production as a function of the left- and right-handed smuons, $\tilde{\mu}_{L,R}$, mass and lifetime at 95\% CLs, for the Displaced Slepton Signature (Phys. Rev. Lett. 127 051802 (2021)~\cite{SUSY-2018-14}), the Intermediate Slepton Signature (the result of this search in this paper) and the Prompt Slepton Signature (Eur. Phys. J. C 80 (2020) 123~\cite{SUSY-2018-32}) reinterpretation. The lines below the graphs show the expected and observed limits from the prompt search where the smuons are assumed to be prompt.
}
\label{fig:excl_limit1}
\end{figure}


\FloatBarrier

\section{Conclusion}

A search has been presented for pairs of opposite electrically charged muons with impact parameters in the millimeter range using 139~\ifb\ of $\sqrt{s} = 13$~\TeV\ \pp\ collision data from the ATLAS detector. This search addresses a gap in coverage of possible new physics signatures between existing searches for leptons with large displacement and prompt leptons. Results are consistent with the SM background prediction. This search provides unique sensitivity to long-lived scalar supersymmetric muon-partners (smuons) with much lower lifetimes than previously targeted by ATLAS searches. Smuon lifetimes down to 1~ps and smuon masses up to 520 GeV are excluded at 95\% confidence level.
 
As no requirements are made on missing energy, displaced vertices, or jets, this result is model-independent and applicable to any BSM model producing at least two opposite sign, isolated displaced muons with transverse momenta greater than 20 GeV.


\section*{Acknowledgements}


We thank CERN for the very successful operation of the LHC, as well as the
support staff from our institutions without whom ATLAS could not be
operated efficiently.
 
We acknowledge the support of
ANPCyT, Argentina;
YerPhI, Armenia;
ARC, Australia;
BMWFW and FWF, Austria;
ANAS, Azerbaijan;
CNPq and FAPESP, Brazil;
NSERC, NRC and CFI, Canada;
CERN;
ANID, Chile;
CAS, MOST and NSFC, China;
Minciencias, Colombia;
MEYS CR, Czech Republic;
DNRF and DNSRC, Denmark;
IN2P3-CNRS and CEA-DRF/IRFU, France;
SRNSFG, Georgia;
BMBF, HGF and MPG, Germany;
GSRI, Greece;
RGC and Hong Kong SAR, China;
ISF and Benoziyo Center, Israel;
INFN, Italy;
MEXT and JSPS, Japan;
CNRST, Morocco;
NWO, Netherlands;
RCN, Norway;
MEiN, Poland;
FCT, Portugal;
MNE/IFA, Romania;
MESTD, Serbia;
MSSR, Slovakia;
ARRS and MIZ\v{S}, Slovenia;
DSI/NRF, South Africa;
MICINN, Spain;
SRC and Wallenberg Foundation, Sweden;
SERI, SNSF and Cantons of Bern and Geneva, Switzerland;
MOST, Taiwan;
TENMAK, T\"urkiye;
STFC, United Kingdom;
DOE and NSF, United States of America.
In addition, individual groups and members have received support from
BCKDF, CANARIE, Compute Canada and CRC, Canada;
PRIMUS 21/SCI/017 and UNCE SCI/013, Czech Republic;
COST, ERC, ERDF, Horizon 2020 and Marie Sk{\l}odowska-Curie Actions, European Union;
Investissements d'Avenir Labex, Investissements d'Avenir Idex and ANR, France;
DFG and AvH Foundation, Germany;
Herakleitos, Thales and Aristeia programmes co-financed by EU-ESF and the Greek NSRF, Greece;
BSF-NSF and MINERVA, Israel;
Norwegian Financial Mechanism 2014-2021, Norway;
NCN and NAWA, Poland;
La Caixa Banking Foundation, CERCA Programme Generalitat de Catalunya and PROMETEO and GenT Programmes Generalitat Valenciana, Spain;
G\"{o}ran Gustafssons Stiftelse, Sweden;
The Royal Society and Leverhulme Trust, United Kingdom.
 
The crucial computing support from all WLCG partners is acknowledged gratefully, in particular from CERN, the ATLAS Tier-1 facilities at TRIUMF (Canada), NDGF (Denmark, Norway, Sweden), CC-IN2P3 (France), KIT/GridKA (Germany), INFN-CNAF (Italy), NL-T1 (Netherlands), PIC (Spain), ASGC (Taiwan), RAL (UK) and BNL (USA), the Tier-2 facilities worldwide and large non-WLCG resource providers. Major contributors of computing resources are listed in Ref.~\cite{ATL-SOFT-PUB-2021-003}.


\clearpage
 
\printbibliography
 
\clearpage
 
\begin{flushleft}
\hypersetup{urlcolor=black}
{\Large The ATLAS Collaboration}

\bigskip

\AtlasOrcid[0000-0002-6665-4934]{G.~Aad}$^\textrm{\scriptsize 103}$,
\AtlasOrcid[0000-0002-5888-2734]{B.~Abbott}$^\textrm{\scriptsize 121}$,
\AtlasOrcid[0000-0002-1002-1652]{K.~Abeling}$^\textrm{\scriptsize 55}$,
\AtlasOrcid[0000-0001-5763-2760]{N.J.~Abicht}$^\textrm{\scriptsize 49}$,
\AtlasOrcid[0000-0002-8496-9294]{S.H.~Abidi}$^\textrm{\scriptsize 29}$,
\AtlasOrcid[0000-0002-9987-2292]{A.~Aboulhorma}$^\textrm{\scriptsize 35e}$,
\AtlasOrcid[0000-0001-5329-6640]{H.~Abramowicz}$^\textrm{\scriptsize 152}$,
\AtlasOrcid[0000-0002-1599-2896]{H.~Abreu}$^\textrm{\scriptsize 151}$,
\AtlasOrcid[0000-0003-0403-3697]{Y.~Abulaiti}$^\textrm{\scriptsize 118}$,
\AtlasOrcid[0000-0003-0762-7204]{A.C.~Abusleme~Hoffman}$^\textrm{\scriptsize 138a}$
% ...
\bigskip
\\

% ...
$^{29}$Physics Department, Brookhaven National Laboratory, Upton NY; United States of America.\\
% ...
$^{35}$$^{(a)}$Facult\'e des Sciences Ain Chock, R\'eseau Universitaire de Physique des Hautes Energies - Universit\'e Hassan II, Casablanca;$^{(b)}$Facult\'{e} des Sciences, Universit\'{e} Ibn-Tofail, K\'{e}nitra;$^{(c)}$Facult\'e des Sciences Semlalia, Universit\'e Cadi Ayyad, LPHEA-Marrakech;$^{(d)}$LPMR, Facult\'e des Sciences, Universit\'e Mohamed Premier, Oujda;$^{(e)}$Facult\'e des sciences, Universit\'e Mohammed V, Rabat;$^{(f)}$Institute of Applied Physics, Mohammed VI Polytechnic University, Ben Guerir; Morocco.\\
% ...
$^{49}$Fakult\"{a}t Physik , Technische Universit{\"a}t Dortmund, Dortmund; Germany.\\
% ...
$^{55}$II. Physikalisches Institut, Georg-August-Universit\"{a}t G\"ottingen, G\"ottingen; Germany.\\
% ...
$^{103}$CPPM, Aix-Marseille Universit\'e, CNRS/IN2P3, Marseille; France.\\
% ...
$^{118}$Department of Physics, New York University, New York NY; United States of America.\\
% ...
$^{121}$Homer L. Dodge Department of Physics and Astronomy, University of Oklahoma, Norman OK; United States of America.\\
% ...
$^{138}$$^{(a)}$Departamento de F\'isica, Pontificia Universidad Cat\'olica de Chile, Santiago;$^{(b)}$Millennium Institute for Subatomic physics at high energy frontier (SAPHIR), Santiago;$^{(c)}$Instituto de Investigaci\'on Multidisciplinario en Ciencia y Tecnolog\'ia, y Departamento de F\'isica, Universidad de La Serena;$^{(d)}$Universidad Andres Bello, Department of Physics, Santiago;$^{(e)}$Instituto de Alta Investigaci\'on, Universidad de Tarapac\'a, Arica;$^{(f)}$Departamento de F\'isica, Universidad T\'ecnica Federico Santa Mar\'ia, Valpara\'iso; Chile.\\
% ...
$^{151}$Department of Physics, Technion, Israel Institute of Technology, Haifa; Israel.\\
$^{152}$Raymond and Beverly Sackler School of Physics and Astronomy, Tel Aviv University, Tel Aviv; Israel.\\
% ...
$^{a}$ Also Affiliated with an institute covered by a cooperation agreement with CERN.\\
$^{b}$ Also at An-Najah National University, Nablus; Palestine.\\
$^{c}$ Also at APC, Universit\'e Paris Cit\'e, CNRS/IN2P3, Paris; France.\\
$^{d}$ Also at Borough of Manhattan Community College, City University of New York, New York NY; United States of America.\\
$^{e}$ Also at Center for High Energy Physics, Peking University; China.\\
$^{f}$ Also at Center for Interdisciplinary Research and Innovation (CIRI-AUTH), Thessaloniki; Greece.\\
% ...

\end{flushleft}


\end{document}